\documentstyle[12pt]{article}
\author{
        {\bf  Marek PAW\L  OWSKI}\thanks{e-mail : 
pawlowsk@fuw.edu.pl.} \\
        {\it  Soltan Institute for Nuclear Studies}\\
        {\it  Warsaw, Poland}}

\title{Gauge theory of phase and scale\thanks{Invited talk given at 
international workshop "Macroscopic Electrodynamics II", November 97, Krak\'ow, Poland}  }

\newcommand{\ben}{\begin{enumerate}}
\newcommand{\een}{\end{enumerate}}
\newcommand{\be}{\begin{equation}}
\newcommand{\ee}{\end{equation}}
\newcommand{\bea}{\begin{eqnarray}}
\newcommand{\eea}{\end{eqnarray}}
\newcommand{\bc}{\begin{center}}
\newcommand{\ec}{\end{center}}

\begin{document}
\maketitle

\begin{abstract}
The old Weyl's idea of scale recalibration freedom and 
the Infeld's and van der Waerden's (IW) ideas concerning 
geometrical interpretation of the natural spinor phase gauge 
symmetry are discussed in the context of modern models 
of fundamental particle interactions. It is argued that IW 
gauge symmetry can be naturaly identified with $U(1)$ 
symmetry of Weinberg-Salam model. It is also argued that 
there are no serious reasons to reject Weyl's gauge theory 
from the considerations. Its inclusion enriches the original 
Weinberg-Salam theory and leads to prediction of new 
phenomena that do not contradict experiments. 

\end{abstract}

\section{Introduction}

Gauge theories are fundamental tools in contemporary 
physics of particles and their interactions. Standard Model 
of fundamental interactions (SM) that reasonable describes 
particle physics at present accelerator energies is a quantum 
gauge theory of $U(1)\times SU(2) \times SU(3)$ symmetry 
group of electroweak and strong forces. The model leads to 
cosmological scenarios that seems to be consistent with 
observational astrophysics. There are many extensions 
and modifications of SM. The scheme of gauge theory is 
a base of all of them. These gauge theories are in fact gauge 
theories of generalized phase of spinorial field multiplets. 
All of them are formulated in flat space time but it is supposed 
(and sometimes it is proved) that they can be generalized (at 
least locally) to a case of arbitrary Riemann space. 

The first consistent formulation of $U(1)$ gauge theory 
of spinor phase in curved space was given soon after Dirac's 
theory was proposed. This model is reviewed shortly in 
section 2. The notion of gauge symmetry is even older. 
It was introduced by Weyl before the notion of spinors has 
been defined. Today we can call this theory a gauge theory 
of scale. Its short review is presented in section 3. Both early 
gauge theories of phase and scale were based on abelian gauge 
groups and was to incorporate electromagnetism into the 
geometrical scheme of general relativity. Those attempts were 
admitted to be unsuccessful. The reasons and arguments for that 
are shortly reviewed in section 4 where a critical discussion of 
hose arguments is also given. In section 5 a general model with 
scale and phase gauge symmetry is described. Its features and 
physical consequences are discussed in section 6.

\section{Gauge theory of phase in Infeld and van der Waerden formulation.}

Soon after the appearance of Dirac's theory of quantum relativistic 
electron in the flat space \cite{dirac28} the general relativistic 
extension of Dirac's theory was also proposed 
\cite{schouten,schroedinger,infeld}. A canvas for such description 
is a four dimensional manifold M. A copy of two dimensional 
complex vector field $F_pM$ is attached to every point $p$ of M. 
Two, in principle independent pairs of affine and metric structures 
can be implemented on M. The tangent boundle $TM$ can be 
equipped with an affine connection $\Gamma$ and the field of 
metric $g$. Independently a connection $\gamma$ can be 
defined in the boundle $FM$. For generic two dimensional c
omplex vector space there is a natural class of antisymmetric 
Levi-Civita metrics that differ by a complex factor. Thus arbitrary 
field $\varepsilon$ of Levi-Civita mertic can be chosen at $FM$. 
The important observation is that the Levi-Civita metric $\varepsilon$ 
induce Lotentz metric $\varepsilon\otimes\overline{\varepsilon}$ at 
every fiber of $FM\otimes \overline{FM}$ (see e.g. \cite{wald} for 
definition of complex conjugation structure "$\bar{~}$" and for 
further details). Thus the real part of $FM\otimes \overline{FM}$ 
(which is four dimensional real vector boundle; let us denote it 
$F\bar F M$) can be naturally related with $TM$ - the tangent 
vector boundle of M.

In Einstein's general relativity theory the affine and metric structures 
of $TM$ are related by metricity condition 
\be 
\nabla g=0
\label{metricity}
\ee
and torsion free condition
\be
\Gamma^{\lambda}_{\mu\nu}-\Gamma^{\lambda}_{\nu\mu}=T^{\lambda}_{\mu\nu}=0.
\label{torsion}
\ee

Keeping those restrictions and relating $TM$ with $F\bar F M$ 
Infeld and van der Waerden \cite{infeld} found that metric 
structure $\varepsilon$ of $FM$ is given by metric structure 
$g$ of $TM$ up to the arbitrary phase factor while the affine 
structure $\gamma$ of $FM$ is given by the affine structure 
$\Gamma$ of $TM$ up to an arbitrary vector field. It is clear 
that this new field is a compensating potential for the $U(1)$ 
local symmetry group of phase transformations of all Dirac 
fields in the theory. The authors have identified this new field 
with electromagnetic potential.

\section{Gauge theory of scale: conformal Weyl's model.}

The Infeld and van der Waerden model wasn't the first example 
of gauge theory. The idea and notion of gauge invariance was 
introduced by Weyl \cite{weyl} as a consequence of natural 
generalization of Riemann geometry. Weyl assumed that the 
metricity condition (\ref{metricity}) can be replaced by a less 
restrictive condition
\be
\nabla g_{\mu\nu}\sim g_{\mu\nu}.
\label{weyl}
\ee
Thus he supposed that for a vector transported around a closed 
loop by parallel displacement not only the direction but also the 
length can change but the angle between two parallelly 
transported vectors has to be conserved. 

If the Einstein's torsion free condition (\ref{torsion}) is kept 
then again there is a relation between the metric and the affine 
structures of $TM$ but now the connection is not given uniquely 
by the Christophel symbol. It depends also on an arbitrary 
vector field. This field is the compensating potential for the 
local gauge group of length changes of all dimensional fields 
in the theory. Originally Weyl interpreted this new field as 
electromagnetic vector potential. Soon he abandoned both the 
electromagnetic interpretation and the whole idea that his 
new symmetry (called the conformal symmetry as it conserves 
angles) plays any role in physics.

\section{Abandoned models.}

Both models - the Weyl's gauge theory of scale and the Infeld 
and van der Waerden gauge theory of phase - despite their 
geometrical beauty, was abandoned for physical reasons. 

Weyl's theory that leaves the freedom for the space-time 
dependent choice of length standards was rejected on the 
base of argument that it clashes with quantum phenomena 
that provide an absolute standard of length. Thus, at least, 
there is no need for the arbitrary metric standards of Weyl's 
theory. On the other hand the electromagnetic interpretation 
of this theory seemed to be not satisfactory by itself (see 
however \cite{dirac73}). 

The physical reasons for rejection of Infeld and van der Waerden 
interpretation of their vector potential as a medium for 
electromagnetic interaction was the fact that the obtained 
potential couples universally to all fermions. Thus it should 
couple also to neutrinos that are electrically neutral Dirac 
particles. Consequently the $U(1)$ gauge symmetry of 
fermion phase should be considered as a possible new 
independent gauge symmetry. As there is no other long 
range interactions observed in nature except electromagnetism 
and gravity, it is assumed that such a gauge interaction is 
not realized or is extremely small.

Let us revise critically all the arguments mentioned above. 

There were two kinds of arguments against Weyl's theory. 
The first of them laid down the law a contradiction between 
the theory and quantum phenomena. Those convictions are 
mostly based on a misunderstanding or misinterpretation of 
Weyl's gauge symmetry. In fact the freedom to set arbitrary 
length standards along an atomic path does not mean that 
atomic frequencies will depend on the atomic histories what
 was the most popular argument in early literature. In Weyl's 
theory atomic frequency depends on the length standard at a 
given point but simultaneously all other dimensional quantities 
measured at this point depend on this standard in the same way. 
There is no contradiction with experiment as dimensionless 
ratios are standard independent and of course do not depend 
on the history of a particular atom. 

More serious arguments against Weyl's theory were based on 
the reasonable claim that the acceptable theory should not 
introduce needles objects and notions. If atomic clocks 
measure time in an absolute way and velocity of light is an 
absolute (or at least definite) physical quantity then the 
relativism of length is unnatural and redundant. It hears 
very reasonable except one subtle question: what atomic 
clock provides the absolute time and length standard? The 
fast answer is: ALL! But here the further problems begin. 
However we know form our "almost flat" experience that 
"free atomic clock" frequency ratios are external conditions 
independent, should we really extrapolate those experience 
to all conditions and times? A naive extrapolation could be 
evidently wrong as we know from solid state physics. We 
can imagine very strong sources of gravity producing such 
extremal conditions that nor known atomic or quantum clocks 
will exist there. And what about the radiation age of Universe 
when there (here) were no matter at all? Observe that nor tricks 
with so called "distant" or "isolated" standards are helpful in 
the case of gravity as there is no screening of interactions. 
Of course we are free to assume that - roughly speaking - 
the ratios of electron mass to proton mass and to other 
quantum standards are always and everywhere the same 
but we should remember (especially when we interpret 
such effects like red shift or other distant signals) that this 
is only our assumption and it could be and it should be a 
subject of experimental verification. Weyl's theory makes 
a room to relax from such at least not definitely confirmed 
suppositions. Can we judge {\sl a priori} that it is really needless?

\smallskip

The arguments formulated against Infeld's and van der 
Waerden's interpretation of vector gauge potential (the 
potential that arises when the affine structure of tangent 
boundle is extended to spinor boundle) are based on the
 fact that neutrinos are chargeless. Those arguments were 
important before Weinberg-Salam theory (WS) has been 
proposed. WS predicts that all fermions couple to $U(1)$ 
gauge field. There is a second nonabelian gauge group 
$SU(2)$ in the theory acting only on left components of 
Dirac bispinors. Due to the structure of couplings and the 
effective mass matrix for gauge bosons the massless field - 
naturally identified with photon - is a combination of original 
$U(1)$ and $SU(2)$ bosons. It does not couple to neutrinos 
despite the fact that the original abelian vector potential does. 
Thus we are free to identify Infeld - van der Waerden 
potential with $U(1)$ gauge group potential of the WS 
model without any conflict with theory and experiment.

\smallskip

We see that the arguments raised against Weyl's and Infeld - 
van der Waerden models are not ultimate and definitively 
convicting. On the other hand both theories realize in a 
sense an old and beautiful idea that physical interactions 
should be ascribed to geometrical properties of space itself, 
instead of being merely something embedded in space. 
Observe that both these theories are complementary and 
correlated. Weyl potential can be raised to spinorial level 
according to Infeld and van der Waerden prescription. 
Then it can be collected (together with the derivative of  
$\log{|\det{\varepsilon}|}$) to be the real part of a complex 
vector potential that has an imaginary part found by Infeld 
and van der Waerden \cite{hayashi}. The Infeld - van der 
Waerden correlation between geometrical structures of 
$TM$ and $FM$ leads immediately to Weyl's conformal 
metricity condition (\ref{weyl}). This result is independent 
on any assumption about relation between metrical and affine s
tructures of $TM$ (e.g. is independent on (\ref{torsion})) and 
follows only from the fact that metric $g$ is related with 
spinorial metric $\varepsilon$ by an arbitrary Infeld - van 
der Waerden relation which, for selfconsistency of the model, 
has to be covariantly constant. Thus the correlations between 
gauge theories of phase and scale are rich and universal.

\section{Classical gauge theory model of phase and scale.}

Despite the controversies around  geometrical origin of  $U(1)$ g
auge theory of fermion phase its role in physics is out of dispute. 
The conformal gauge theory of scale is less lucky but many 
authors returns to the original Weyl's ideas in various contexts 
(se e.g. \cite{hehl} and also \cite{dirac73,padman,cheng,foundations}).
Let us write down a general model respecting both those 
symmetries. But first let us fix the notation.

Weyl's potential will be denoted by $S_{\mu}$.  Then, if torsion 
free condition (\ref{torsion}) is assumed, the connection is given by
\be
\Gamma^{\rho}_{\mu\nu}=\{^{\rho}_{\mu\nu}\}+
f(S_{mu}g^{\rho}_{\nu}+S_{\nu}g^{\rho}_{\mu}-S^{\rho}g_{\mu\nu})
\label{connection}
\ee
where $f$ is an arbitrary coupling constant  (in principle it could 
be absorbed at this level by a redefinition of $S_{\mu}$ but it is 
convenient to keep it here and set its value later). 
Consequently
\be
\nabla_{\mu} \hat{g} = -2fS_{\mu}\hat{g}.
\label{weyl2}
\ee
Equations (\ref{connection}) and (\ref{weyl2}) are invariant with 
respect to Weyl transformations 
\be
g_{\mu\nu}\to \Omega^2 g_{\mu\nu}=e^{2\lambda} g_{\mu\nu}
\label{transmetric}
\ee
\be
S_{\mu}\to S_{\mu}-{1\over f}\partial_{\mu}\lambda.
\label{transweyl}
\ee
Thus metric tensor is covariant with respect to Weyl transformations 
with degree 2. The Riemann and Ricci tensors constructed from 
(\ref{connection}) are conformally invariant objects but the scalar 
curvature $R$ is not. $R$ can enter linearly to a conformally 
invariant expression of dimension of action if it is combined 
with a scalar field  $\phi$ that transforms according to 
\be
\phi \to e^{-\lambda}\phi.
\label{transphi}
\ee
Then the combination $\phi^2R$ is conformally invariant. The 
conformal covariant derivative of $\phi$ is given by
\be
\nabla_{\mu}\phi = (\partial_mu -fS_{\mu})\phi
\label{nablaphi}
\ee
and it transforms according to (\ref{transphi}).

The most general conformally invariant lagrangian that leads 
to second order equations of motion for the metric-Weyl-scalar system reads \cite{padman}:
\be
L_g= -{\alpha_1\over 12}\phi^2 R + {\alpha_2\over 2}\nabla_{\mu}
\phi\nabla^{\mu}\phi - {\alpha_3\over 4} 
H_{\mu\nu}H^{\mu\nu} -{\lambda\over 4{\rm !}}\phi^4
\label{Lg}
\ee
where
\be
H_{\mu\nu}=\partial_{\mu}S_{\nu}-\partial_{\nu}S_{\mu}.
\label{Hmunu}
\ee
The coupling constants $\alpha_1$, $\alpha_2$ and $\alpha_3$ are 
arbitrary but the last two can be absorbed in $\phi$ and $\S_{\mu}$ 
by a suitable redefinition of the fields. Observe however, that we are 
not able to absorb simultaneously $\alpha_3$ and $f$. Thus the last 
coupling remains arbitrary and has to be fixed by experiment.

Now we can include fermions. First we should recall \cite{hayashi} 
that Weyl's vector potential $S_{\mu}$ do not couple directly to Dirac 
fermions if they transforms according to the rule
\be
\Psi\to e^{-{3\over 2}}\Psi.
\label{transpsi}
\ee
If we want to fit to the SM prescription we have to admit that except 
for the $U(1)$ gauge symmetry group of fermion phase - let the 
gauge potential of it be denoted by $B_{\mu}$ - also other internal 
nonabelian gauge symmetry groups are present in the model. The 
scalar field $\phi$ that has been introduced in (\ref{Lg}) can be 
extended to a complex scalar multiplet. The ordinary derivative 
in (\ref{nablaphi}) must be replaced by $D_{\mu}= \nabla_{\mu} - 
ieB_{\mu} + ...$ being the convariant derivative with respect 
to $U(1)$ (we assume that it couples universally  to the phase 
of  $\phi$) and with respect to some other internal symmetry groups
\be
\nabla_{\mu}\phi = (D_{\mu} -fS_{\mu})\phi.
\label{nablaphi2}
\ee
Thus the curvilinear versions of  Dirac lagrangian $L_{\Psi}$ and 
Yukawa lagrangian $L_Y$ can be easily written. We can also select 
Maxwell lagrangian $L_B = -{1 \over 4} F_{\mu\nu}F^{\mu\nu}$ 
for $U(1)$ vector potential and a general Yang-Mills lagrangian 
$L_V$ for other gauge potentials. As there are two abelian gauge
 groups in the model also a mixed term
\be
L_{SB}=\alpha_4 H_{\mu\nu}F^{\mu\nu}
\label{LSB}
\ee
respects all symmetries and has to be admitted. 
The total lagrangiam can be written as a sum of  these terms
\be
L_T=L_g+L_{\Psi}+L_Y + L_B +L_V + L_{SB}.
\label{LT}
\ee

\section{Discussion.}

The theory given by (\ref{LT}) has interesting properties 
that depend on the value of coupling constants $\alpha_i$. 
But its special property is its conformal gauge invariance. 
If a gauge theory is to be solved some additional gauge fixing 
conditions have to be supposed  in order to make the 
evolution definite. This choice is arbitrary within the whole 
class of gauge equivalent conditions. The physical results 
are gauge choice independent. There are custom procedures 
to handle this freedom in the case of gauge symmetry of 
generalized phase. The case of scale gauge symmetry is 
specially interesting. The reference dimensional scale can 
be chosen arbitrary but it is reasonable to choose it in a way 
that is most practical and convenient. If we are focused on 
laboratory phenomena where gravitational effects are negligible 
then there is no reason to doubt in universality of length 
standards provided by the whole class of quantum phenomena 
please recall the discussion of section 4). We are free to chose 
the length standards that lead to constant, space independent 
particle masses. If the theory (\ref{LT}) is a conformal 
modification of SM then the conformal gauge fixing condition 
that provides correspondence with the ordinary description is 
the condition \cite{cheng,foundations}
\be
|\phi|^2=v^2=(246GeV)^2.
\label{fermi}
\ee
It leads to the mass spectrum that is the same as obtained from 
the mechanism of spontaneous symmetry breaking in WS 
but those mechanism is absent in the minimal version of 
conformal theory. As a result also Weyl's vector field $S_{\mu}$ acquires mass
\be
m_S^2={1\over 2}f(\alpha_2-\alpha_1)v^2
\label{ms}
\ee
that is equal  zero only in the special case when $\alpha_2=\alpha_1$ 
and an additional symmetry is realized in the model.

The striking feature of the described theory is the lack of 
ordinary Einstein term in (\ref{LT}). Observe however, 
that with the condition (\ref{fermi}) this term can be easily 
reproduced \cite{cheng}. It is sufficient to demand that  
\be
-{\alpha_1\over 12}v^2={1\over 8\pi G}.
\label{newton}
\ee
It leads to Weyl vector mass
\be
m_S=0.5\cdot 10^{19} f\cdot GeV.
\ee

It was already mentioned that in the case  $\alpha_2=\alpha_1$ 
the model has an additional symmetry. The Weyl potential 
decouples from scalar field and if $\alpha_4=0$ it is coupled 
only to gravity. Transformations (\ref{transmetric}), 
(\ref{transphi}) and (\ref{transpsi}) are the symmetries of 
the theory independently on (\ref{transweyl}). We get 
Penrose-Chernikov-Tagirov theory of scalar field conformally 
coupled with gravity
\cite{penrose}. We are free to include further terms to (\ref{LT}) 
that respects the new symmetry. Thus, despite the fact that the 
coefficient standing in front of  $R$ in the original lagrangian  
is negative we are able to reproduce appropriate Newtonian limit 
of the whole theory \cite{foundations,part1}.

The very new feature of lagarngian (\ref{LT}) is the mixed term 
(\ref{LSB}) that leads to interaction of Weyl and $U(1)$ vector 
potentials. At quantum level  it would result in a mixing of Weyl 
boson with photon and weak bosons - the effect in a sense similar 
to the known $\gamma - Z$ mixing. As the mass of $S_{\mu}$ 
and the coupling $\alpha_4$ is not predicted by the theory the 
strength of the mixing effect could be small as well as very large. 
Also the mass $m_S$ cannot be easily estimated from the known 
data as there is no interaction of fermions with the Weyl potential. 
Thus definite answers concerning the presence and interactions of 
Weyl potential should be looked in experiments.

{\bf Acknowledgments}

The author is indebted to prof. E. Kapu\'scik, prof. N. Konopleva  
and dr A. Horzela for valuable discussion. I'm also grateful  to 
prof. I. Bia\l ynicki-Birula for rendering a copy of English translation 
of \cite{infeld}. The work was supported by Polish Committee for 
Scientific Researches grant no. 2 P03B 183 10.
.


\begin{thebibliography}{99}
\bibitem{dirac28} P.A.M. Dirac, Proc. Roy. Soc. (London), {\bf A117}, 610 (1928).
\bibitem{schouten} J.A. Schouten, Journal of Math. and Phys., {\bf 10}, 239 (1931).
\bibitem{schroedinger} E. Schroedinger, S.-B. Preuss. Akad. Wiss., 105 (1932).
\bibitem{infeld} L. Infeld and B.L. van der Waerden, S.-B. Preuss. Akad. Wiss., 380 (1932).
\bibitem{wald} R.M. Wald, {\sl General Relativity}, The University of Chcago Press, 1984.
\bibitem{weyl} H. Weyl, S.-B. Preuss. Akad. Wiss., 465 (1918); Math. Z. {\bf 2}, 
384 (1918); Ann. der Phys. {\bf 59}, 101 (1919).
\bibitem{dirac73} P.A.M. Dirac, Proc. Roy. Soc. (London) {\bf A333}, 403 (1973).
\bibitem{hayashi} K. Hayashi and T. Shirafuji, Prog. Theor. Phys. {\bf 57}, 302 (1977); 
K. Hayashi, M. Kasuya and T. Shirafuji, {\sl ibid}, p. 431.
\bibitem{hehl} F. W. Hehl, J.D. McCrea, E. W. Mielke,
Y. Neeman, Phys.Rept. {\bf 258}, 1 (1995). 
\bibitem{padman} T. Padmanabhan, Class. Quantum Grav. {\bf 2}, L105 (1985).
\bibitem{cheng} H. Cheng, Phys. Rev. Lett. {\bf 61}, 2182 (1988).
\bibitem{foundations} M. Paw\l owski and R. R\c aczka, Found. Phys. {\bf 24}, 1305 (1994).
\bibitem{penrose} R. Penrose, in {\sl Relativity, Groups and Topology}, 
Gordon and Brach, London 1964; N.A. Chernikov and E.A. Tagirov, Ann.
Inst. H. Poincare {\bf 9}, 109 (1968).
\bibitem{part1} M. Paw\l owsk and R. R\c aczka, {\sl A Higgs-free Model for 
Interactions. Part 1}, preprint SISSA-Trieste, ILAS/EP-3-1995; hep-ph 9503269.

\end{thebibliography}
\end{document}